\documentclass[%
preprint,
superscriptaddress,
showpacs,
preprintnumbers,
graphicx,
floatfix,
byrevtex,
amsfonts,
amsmath,
amssymb,
aps,
prl,
]{revtex4-1}

\usepackage{graphicx}
\usepackage{dcolumn}
\usepackage{bm}
\usepackage[mathlines]{lineno}
\usepackage{ulem}
\usepackage{color}

\begin{document}
\preprint{} 

\title[Short Title]{Ultrafast Dynamics of Electron-phonon Coupling in Transition-metal Dichalcogenides}

\author{Kotaro Makino}
\email{k-makino@aist.go.jp}
\affiliation{Nanoelectronics Research Institute, National Institute of Advanced Industrial Science and Technology (AIST), Tsukuba Central 5, 1-1-1 Higashi, Tsukuba 305-8565, Japan}
\author{Yuta Saito}
\affiliation{Nanoelectronics Research Institute, National Institute of Advanced Industrial Science and Technology (AIST), Tsukuba Central 5, 1-1-1 Higashi, Tsukuba 305-8565, Japan}
\author{Shuuto Horii}
\affiliation{Graduate School of Science and Engineering, Saitama University, Saitama 338-8570, Japan}
\author{Paul Fons}
\affiliation{Nanoelectronics Research Institute, National Institute of Advanced Industrial Science and Technology (AIST), Tsukuba Central 5, 1-1-1 Higashi, Tsukuba 305-8565, Japan}
\author{Alexander V. Kolobov}
\affiliation{Nanoelectronics Research Institute, National Institute of Advanced Industrial Science and Technology (AIST), Tsukuba Central 5, 1-1-1 Higashi, Tsukuba 305-8565, Japan}
\author{Atsushi Ando}
\affiliation{Nanoelectronics Research Institute, National Institute of Advanced Industrial Science and Technology (AIST), Tsukuba Central 5, 1-1-1 Higashi, Tsukuba 305-8565, Japan}
\author{Keiji Ueno}
\affiliation{Graduate School of Science and Engineering, Saitama University, Saitama 338-8570, Japan}
\author{Richarj Mondal}
\affiliation{Division of Applied Physics, Faculty of Pure and Applied Sciences, University of Tsukuba, 1-1-1 Tennodai, Tsukuba 305-8573, Japan}
\author{Muneaki Hase}
\email{mhase@bk.tsukuba.ac.jp}
\affiliation{Division of Applied Physics, Faculty of Pure and Applied Sciences, University of Tsukuba, 1-1-1 Tennodai, Tsukuba 305-8573, Japan}

\date{\today}

\begin{abstract}
Time-domain femtosecond laser spectroscopic measurements of the ultrafast lattice dynamics in $2H$-MoTe$_{2}$ bulk crystals were carried out to understand the 
carrier-phonon interactions that govern electronic transport properties. 
An unusually long lifetime coherent $A_{1g}$ phonon mode was observed even in the presence of very large density of photo-excited carriers at room temperature.
The decay rate was observed to decrease with increasing excitation laser fluence.
Based on the laser fluence dependence including the inducement of significant phonon softening and a peculiar decrease in phonon decay rate, we attribute the long lifetime lattice dynamics to weak anharmonic phonon-phonon coupling and a carrier-density-dependent deformation potential electron-phonon coupling.


\end{abstract}

\maketitle
Transition-metal dichalcogenides (TMDCs) are recognized as one of the most promising two-dimensional (2D) layered materials for optical, mechanical, and electrical applications owing to their unique van der Waals bonded multi-layered structures \cite{Mak10, Jariwala13, Kolobovbook}, the inherent band-gap engineering of topological properties \cite{Qian14, Sun15, Huang16}, and the predicted high carrier mobility \cite{Zhang14}.
Molybdenum ditelluride (MoTe$_{2}$), a representative TMDC material, is an ideal material for optical and electrical applications due to its unique layered 2D structure.
MoTe$_{2}$ can crystallize in three different phases, i.e. the room temperature $2H$, the high-temperature $1T^{\prime}$ and the low-temperature $T_{d}$ phases \cite{Keum15}.
$2H$ phase has a hexagonal structure while the $1T^{\prime}$ phase has a monoclinic structure and transforms into distorted monoclinic $T_{d}$ phase.
The $2H$-MoTe$_{2}$ is a semiconductor and holds great promise for field-effect transistors \cite{Fathipour14, Cho15} and optical applications \cite{Octon16,Yin16}.

A fundamental understanding of the interaction between electrons and the lattice is of central importance for advancing TMDCs to device applications as, in addition to impurity and vacancy scattering, electron-phonon coupling often limits carrier mobility in single crystal materials \cite{Yu01}.
In an electrical devices, electron-phonon coupling plays a significant role in governing device performance.
For example, interactions between phonons and photo-excited carriers are a crucial issue for TMDC-based optoelectronic device applications.
However, little has been experimentally unveiled in terms of electron-phonon coupling in TMDCs although the Fr\"{o}hlich interaction has been theoretically argued to play an important role in electron-phonon coupling in TMDCs \cite{Sohier16}.
Coherent phonon spectroscopy \cite{Dekorsy00} is a powerful experimental tool to explore the lattice degrees of freedom on ultrafast time scales and enables investigation of electron-phonon coupling dynamics as well as ultrafast phase transitions \cite{Rini07, Makino11, Kim12, Makino12}. 

In this Letter, we have investigated the electron-phonon coupling dynamics in a bulk crystal of $2H$-MoTe$_{2}$ by employing optical pump-probe coherent phonon spectroscopy with varying pump fluences ($F$).
The lifetime of the coherent $A_{1g}$ mode was found to be remarkably long in spite of the presence of a large number of photo-excited carriers.
The exploitation of coherent phonon states can be important for quantum computing \cite{Tesch04}, as they have the potential for operation in the terahertz (THz) range since coherent optical phonons typically oscillate in the THz frequency region.
For this purpose, the lifetime of coherent phonons is an important parameter in the retention of data necessary for the realization of complex computing.
In general, however, coherent phonons typically damp out within a few picosecond in most materials \cite{Hase10R}. 
Wide gap semiconductors are an exception that exhibit long-lifetime coherent phonons \cite{Yee02, Lee03, Ishioka06}.
However, in the presence of carrier excitation, these materials exhibit a coherent phonon lifetime that is drastically shortened due to carrier-phonon scattering \cite{Yee02, Ishioka10}. 
Here, we show that MoTe$_{2}$ is an exception even in the presence of a substantial concentration of photo-excited carriers. 

Optical pump-probe measurements were carried out using a femtosecond Ti:sapphire laser oscillator operated at 80 MHz, which provided near infrared optical pulses with a pulse duration of $\leq$ 30 fs and a central wavelength of 800 nm. The average fluence of the pump beam was varied from 50 to 400 $\mu$J/cm$^{2}$.  
The $s$-polarized pump and the $p$-polarized probe beam were co-focused onto the sample to a spot size of $\approx$25 $\mu$m. 
The optical penetration depth at 800 nm was estimated from the absorption coefficient to be $\approx$57 nm. 
The delay between the pump and probe pulses was scanned by an oscillating retroreflector operated at a frequency of 19.5 Hz up to 15 ps \cite{Hase12}. 
The transient reflectivity change ($\Delta R/R$) was recorded as a function of pump-probe time delay. The measurements were performed in air at room temperature.
The sample used was a small flake of $2H$-MoTe$_{2}$ single crystal with the $c$-axis of the crystal corresponding to the sample normal. 

Fig. 1(a) shows a typical time-domain signal observed with $F$ = 50 $\mu$J/cm$^{2}$.
At 0 ps (the arrival of the pump pulse), a clear oscillatory pattern emerges and persists for more than 30 ps.
The amplitude of the oscillatory pattern decreases with time.
Fig. 1(b) shows the Fourier transformed (FT) spectrum of the time domain signal [Fig. 1(a)].
Two distinct peaks can be seen at around 3.6 and 5.1 THz.
As can be seen in Fig. 1(c), $2H$-MoTe$_{2}$ has a multi-layered structure with van der Waals interactions.
The observed 3.6 and 5.1 THz peaks correspond to the transverse optical (TO) $E_{1g}$ and the longitudinal optical (LO) $A_{1g}$ Raman-active phonon modes, respectively.
Since the bulk crystal of $2H$-MoTe$_{2}$ has an indirect band-gap of $\approx$1.0 eV \cite{Grant75, Conan79}, near-infrared (1.55 eV) optical excitation introduces photo-excited carriers in the conduction bands and hence carrier-phonon coupling plays an important role in the decay of the coherent phonons, as schematically depicted in Fig. 1(c), in addition to other phonon scattering mechanisms, such as phonon-phonon, phonon-impurity, and phonon-defect scattering.
Nevertheless, the lifetime of coherent oscillation in $\Delta R/R$ signal is extremely long, lasting for more than 30 ps.
The narrow bandwidth of the $A_{1g}$ mode compared to the $E_{1g}$ as shown in Fig. 1(b) also indicates that the lifetime of the $A_{1g}$ mode is longer than that of the $E_{1g}$ mode.

\begin{figure}[htbp]
\includegraphics[width=85mm]{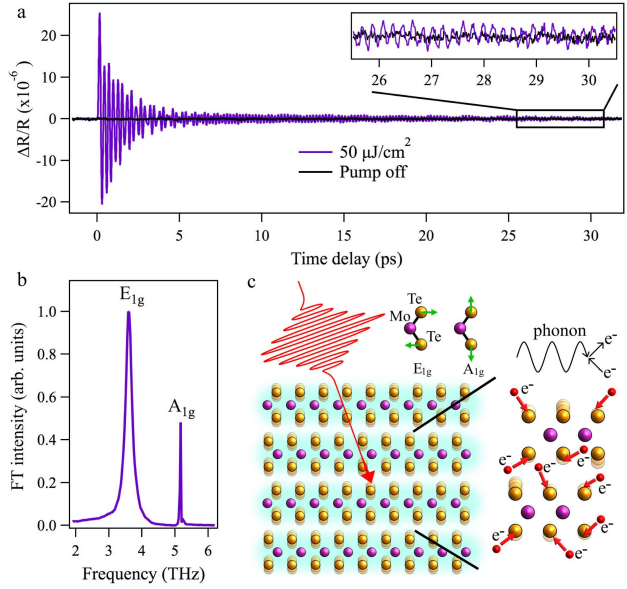}
\caption{(a) Time-domain signal of MoTe$_{2}$ measured with $F$ = 50 $\mu$J/cm$^{2}$. (b) Fourier transformed spectrum of (a). In the spectrum, the $E_{1g}$ and $A_{1g}$ Raman-active coherent phonon modes were observed. (c) The crystal structure and corresponding atomic motion of the coherent phonon modes are illustrated.}
\label{Fig. 1}
\end{figure}

Fig. 2 shows a discrete Wavelet transform (DWT) spectrogram for $F$ = 50 $\mu$J/cm$^{2}$, a time-frequency-domain representation of Fig 1(a) \cite{Hase03}.
Both of the two dominant optical phonon modes can be seen at their corresponding frequencies, and decay with different time scales.
The lifetime of the $E_{1g}$ mode was found to be much shorter than that of the $A_{1g}$ mode, 
and only the $A_{1g}$ mode signal persists beyond $\approx$8 ps up to the limit of the time delay of the measurement in accordance with the difference in the spectral bandwidth of these modes [see Fig. 1(b)].
Thus, the $A_{1g}$ mode exhibited a long lifetime.

\begin{figure}[htbp]
\includegraphics[width=75mm]{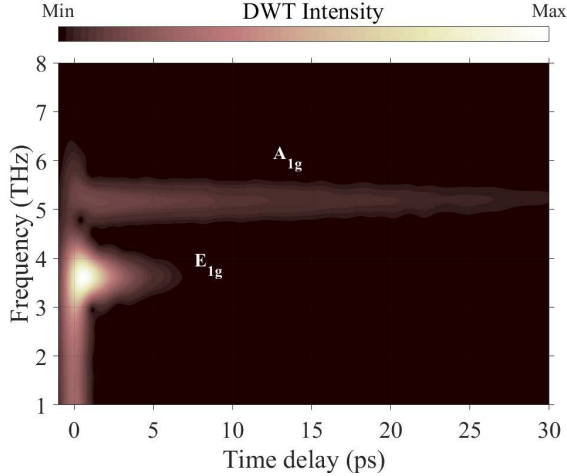}
\caption {Discrete Wavelet transform spectrogram for $F$ = 50 $\mu$J/cm$^{2}$.}
\label{Fig. 2}
\end{figure}
	
Fig. 3 demonstrates the time-domain signals observed for various fluences for early (from -0.5 to 8 ps) and late (from 22 to 31.5 ps) time windows to emphasize the characteristic behaviors of $E_{1g}$ mode and the long-lifetime of the $A_{1g}$ mode.
As shown in Fig. 3(a), it is interesting to note that the initial amplitude of the oscillatory signal increases sublinearly with increasing $F$  in the range of 100 to 400 $\mu$J/cm$^{2}$. 
By comparison with the signal for $F$ = 100 $\mu$J/cm$^{2}$, the initial amplitude of the oscillation for $F$ = 200 $\mu$J/cm$^{2}$ was found to increase by a factor of two, as expected from the ratio of $F$ values.
The initial amplitudes for $F$ = 300 and 400 $\mu$J/cm$^{2}$ exhibit, however, almost same magnitude, implying a saturation effect occurs, similar to the case for the coherent LO mode in CdTe \cite{Ishioka06j}.
As shown in Fig. 3(b), in fact, the decay rate of the coherent phonons increases with increasing pump fluence.

To further investigate the dynamics observed in photo-excited MoTe$_{2}$, we fitted the time-domain signals with a combination of two damped harmonic oscillators,
\begin{eqnarray}
\Delta R(t)/R = A_{E_{1g}}e^{- t/\tau_{E_{1g}}} \cos (\omega_{E_{1g}} t + \varphi_{E_{1g}}) \nonumber \\
+ A_{A_{1g}}e^{- t/\tau_{A_{1g}}} \cos (\omega_{A_{1g}} t + \varphi_{A_{1g}}) + C.
\label{Eq1}
\end{eqnarray}
where $A$, $\omega$, and $\varphi$ are the amplitude, the frequency, and the initial phase, respectively, where the subscripts $E_{1g}$ and $A_{1g}$ represent the phonon modes, and $C$ is a constant background.
As shown in Fig. 3(a), the experimental data can be fit well especially for early times.
For later time windows, there is significant deviation from the fit towards the end of the observed signals, e.g., at 30 ps.
After $\approx$26 ps, the oscillatory periods of the experimental data becomes shorter than the fit, where the frequencies $\omega$ are kept constant. 
This result suggests that phonon hardening (frequency up-chirping) has occurred for the coherent $A_{1g}$ mode within a time window of $\approx$30 ps.

\begin{figure}[htbp]
\includegraphics[width=75mm]{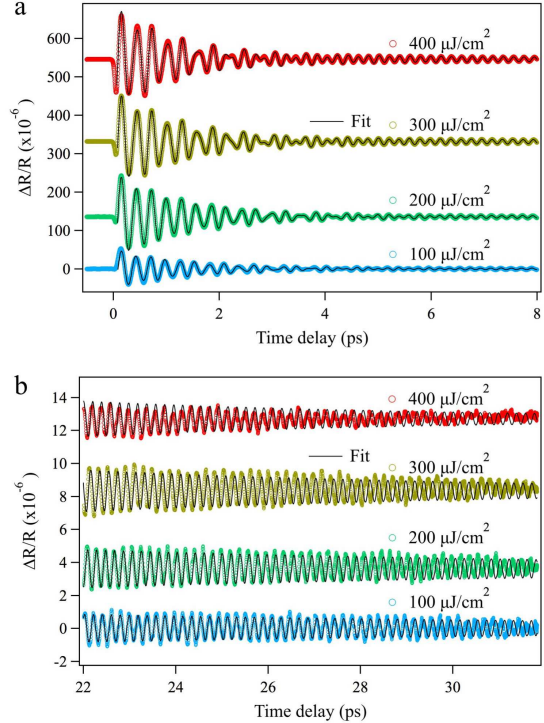}
\caption {Time-domain $\Delta R/R$ signals for $F$ = 100, 200, 300, and 400 $\mu$J/cm$^{2}$ in the time region of -1 $\sim$ 8 ps (a) and 23 $\sim$ 31 ps (b). The colored and the black lines represent the experimental data and the fit using Eq. (1), respectively. Each data set has been vertically offset to avoid overlap.}
\label{Fig. 3}
\end{figure}

The FT spectra obtained from the entire $\Delta R/R$ signals for different $F$ levels (100, 200, 300, and 400 $\mu$J/cm$^{2}$) are shown in Fig. 4.
The frequencies of both the $E_{1g}$ and $A_{1g}$ modes were found to significantly red-shift with increasing $F$, as indicated by the dashed lines.
It should be noted that, while a photo-induced temperature rise to $\approx$670 K can cause a phase transition from $2H$- to $1T'$-MoTe$_{2}$ structure \cite{Cho15}, 
in the present experiment the lattice temperature rise estimated using the two-temperature model \cite{Allen} is $\approx$40 K, implying that the maximum lattice temperature after the photo-excitation is $\approx$340 K.
The value is much lower than the phase transition temperature, and therefore, the possibility of a photo-induced phase transition into $1T'$ phase can be excluded.
Instead of a photo-induced phase transition, we argue that the observed phonon frequency softening is due to changes in the electron-phonon coupling, because of the large photo-excited carrier density as will be discussed later.
We note that the possibility of an electronic-excitation induced phase transition into the $2H^{*}$ phase was predicted \cite{Kolobov16}.
However, the $2H^{*}$ phase is an incommensurate excited state of the $2H$ phase and we consider that the photo-excited sates for each fluence can be treated in the same framework of electron-phonon coupling.

\begin{figure}[htbp]
\includegraphics[width=70mm]{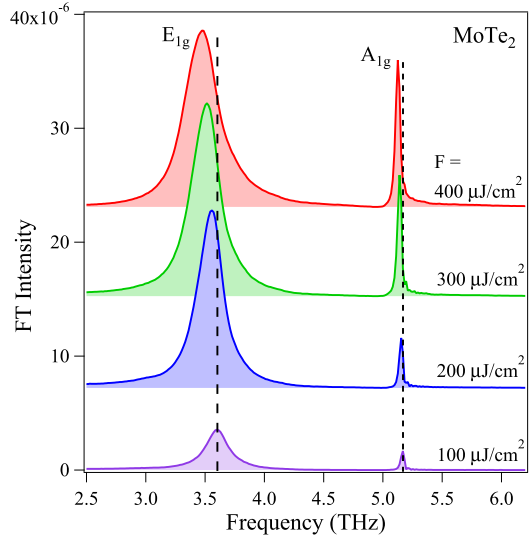}
\caption {FT spectra obtained with $F$ = 100, 200, 300, and 400 $\mu$J/cm$^{2}$. 
The dashed lines represent the peak positions for $F$ = 100 $\mu$J/cm$^{2}$.}
\label{Fig. 4}
\end{figure}

To obtain further insight into the origin of the $F$ dependence of the coherent phonon spectra, the fitting parameters obtained in Fig. 3 for the $E_{1g}$ and $A_{1g}$ modes are plotted as a function of $F$ in Fig. 5.
As $F$ is increased, the initial amplitude of the $A_{1g}$ mode linearly increases, whereas that of the $E_{1g}$ saturates at $\approx$250 $\mu$J/cm$^{2}$ [Fig 5(a)].
This observation might be explained by the effect of the decay on the amplitude \cite{Hase98}, where the $E_{1g}$ mode exhibits a larger decay rate than that of the $A_{1g}$ mode, and/or 
by the saturation of the driving force, e.g., a sublinear carrier density dependent shift of the lattice potential along the $E_{1g}$ coordinate \cite{Katsuki13}. 
The frequencies (decay rates) of both modes monotonically decrease (increase) with increases in $F$.
In terms of the decay rate behavior, the total decay rate (the inverse of the dephasing time, 1/$\tau$) is determined by the sum of the scattering mechanisms, such as phonon-defect scattering due to impurities and vacancies (1/$\tau_{defect}$), anharmonic phonon-phonon coupling (1/$\tau_{anh}$), and electron-phonon coupling (1/$\tau_{e-p}$).
Thus 1/$\tau$ = 1/$\tau_{defect}$+1/$\tau_{anh}$+1/$\tau_{e-p}$.
In a high-quality single crystal, as is the case here, the concentration of impurities and vacancies is quite small so that the $\tau_{defect}$ term is very small and does not depend on $F$.
The anharmonic phonon-phonon coupling depends mainly on the lattice temperature \cite{Klemens66, Vallee94} and the $\tau_{anh}$ is not sensitive to the photo-excited carrier density, although photo-excitation causes a small lattice temperature rise ($\leq$40 K). 
Hence the monotonic increase in the decay rates can predominantly be attributed to the electron-phonon coupling that is also related to phonon softening.

In the scheme of deformation potential (DP) electron-phonon coupling, the frequency and decay rate, i.e., the real and imaginary parts of the phonon self-energy \cite{Cerderia73}, vary in a nearly linear manner with the carrier density.
The estimated number of photo-excited carriers (i.e. 2.1 $\times$ 10$^{19}$ cm$^{-3}$ for $F$= 50 $\mu$J/cm$^{2}$) is much larger than the intrinsic carrier density ($\approx$ 10$^{17}$ cm$^{-3}$) and a change in $F$ is expected to lead to a significant change in the DP interaction.
The red-shift of the frequency can be well modeled by $\Delta\omega \propto -(D_{0}^{2}/\omega_{0})N_{e}$, where $D_{0}$ is the deformation potential, $\omega_{0}$ is the phonon frequency, and $N_{e}$ is the carrier density that is a linear function of $F$ \cite{Cerderia73}.
We can then obtain the ratio of $\Delta\omega_{A_{1g}}/\Delta\omega_{E_{1g}} = (D_{A_{1g}}/D_{E_{1g}})^{2}(\omega_{E_{1g}}/\omega_{A_{1g}})$.
From the linear fit shown in Fig. 5(b), $\Delta\omega_{A_{1g}}/\Delta\omega_{E_{1g}}$ is found to be $\approx$ 0.32. Setting $\omega_{E_{1g}}/\omega_{A_{1g}}$ $\approx$ 0.70, $(D_{A_{1g}}/D_{E_{1g}})^{2}$ = 0.46, and thus $D_{A_{1g}}/D_{E_{1g}}$ = 0.68.
The change in the decay rate, on the other hand, can be expressed by $\Delta\gamma \propto -(D_{0}^{2}\omega_{0})N_{e}$ \cite{Cerderia73}.
From this we obtain the ratio of $\Delta\gamma_{A_{1g}}/\Delta\gamma_{E_{1g}} = (D_{A_{1g}}/D_{E_{1g}})^{2}(\omega_{A_{1g}}/\omega_{E_{1g}})$. 
From the linear fit shown in Fig. 5(c), the ratio $\Delta\gamma_{A_{1g}}/\Delta\gamma_{E_{1g}}$ is found to be $\approx$0.37.
Setting $\omega_{A_{1g}}/\omega_{E_{1g}}$ $\approx$ 1.42, $(D_{A_{1g}}/D_{E_{1g}})^{2}$ = 0.26, $D_{A_{1g}}/D_{E_{1g}}$ = 0.51, a value comparable to the value obtained from the frequency red-shift ($D_{A_{1g}}/D_{E_{1g}}$ = 0.68). 
Given the result that the $D_{A_{1g}}/D_{E_{1g}}$ ratio obtained from a comparison of the decay rate trend is smaller than that derived from the frequency trend, other symmetry- and carrier-density-dependent decay processes, such as the Fr\"{o}hlich interaction, are thought to be not as important as the DP electron-phonon coupling in the current experiment.

The larger DP interaction associated with the $E_{1g}$ mode ($D_{E_{1g}} > D_{A_{1g}}$) explains the stronger damping and larger fluence dependence of the frequency shift and the larger decay rate of the $E_{1g}$ mode over the $A_{1g}$ mode [Fig. 5(b) and (c)]. 
Although we estimate the magnitude ratio of the DP interactions for both modes to be 0.51$^{-1}$ $\approx$ 2.0 at the most, it is unclear why the lifetime of the $A_{1g}$ mode (18.7 ps, i.e., 1/$\tau$ $\approx$ 0.053 ps$^{-1}$) is an order of magnitude larger than that of the $E_{1g}$ mode (1.42 ps, i.e., 1/$\tau$ $\approx$ 0.70 ps$^{-1}$) observed in the lowest fluence limit of $F$ = 50 $\mu$J/cm$^{2}$.
Another important factor might be anharmonic phonon-phonon coupling which is thought to be very small for the $A_{1g}$ mode as observed by the previous Raman measurements which demonstrated the phonon spectral width (the inverse of the decay rate) is insensitive to the lattice temperature \cite{Li16}.
Therefore, $1/\tau_{anh}$ is small for the $A_{1g}$ mode.
When $F$ is small, $1/\tau_{e-h}$ is small and, as a result, $1/\tau$ is small leading the lifetime of the $A_{1g}$ mode to be very long compared with other materials.
Note that the ratio of the $A_{1g}$ mode decay rate for lowest and highest $F$ values is $\approx$1.8, an unusually large value even though the absolute value of the change is small.
This is consistent with the assumption that the DP electron-phonon coupling is the dominant scattering factor for the $A_{1g}$ mode over other interactions.

\begin{figure}[htbp]
\includegraphics[width=75mm]{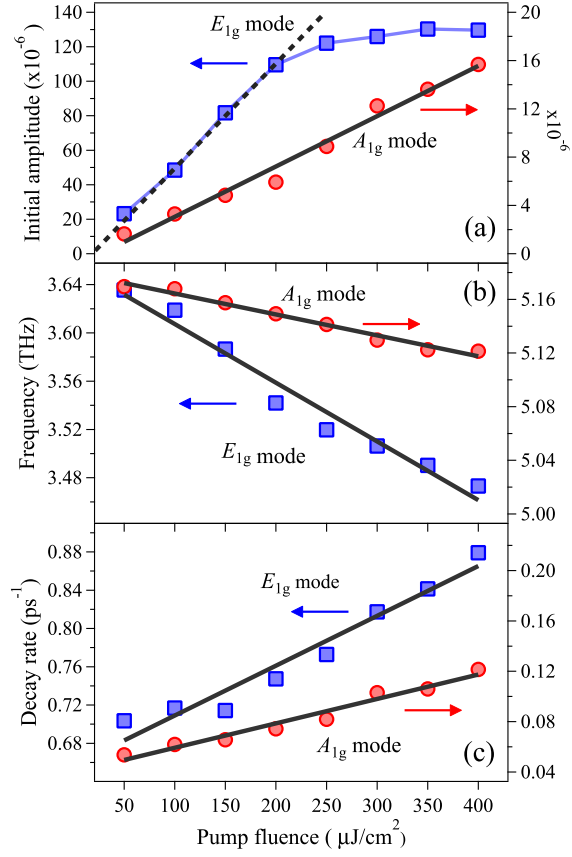}
\caption {(a) The initial amplitudes, (b) the oscillation frequencies, and (c) the decay times of the $E_{1g}$ and $A_{1g}$ modes obtained by fitting of the time-domain signals. The solid lines indicate a linear curve fit. The dotted line serves as a guide to the eye.} 
\label{Fig. 5}
\end{figure}

It is worth mentioning that the $A_{1g}$ lifetime is expected to be insensitive to the number of layers based upon previous report, in which the spectral widths of the Raman peaks for both $E_{1g}$ and $A_{1g}$ modes did not change with an increase in the number of layers, although the peak positions did shift depending on the number of layers \cite{Froehlicher15, Goldstein16}.
In addition, the $A_{1g}$ mode lifetime at room temperature is expected to be comparable to that at low temperature \cite{Li16}.
Therefore, to exploit the long-lifetime of the coherent $A_{1g}$ mode, neither a preparation of an atomically thin layer or low temperature is necessary.
In other words, MoTe$_{2}$ is robust in terms of changes in the environment and expected to provide stable performance for practical phonon-based applications.
In terms of optoelectrical applications, MoTe$_{2}$ is expected to pave the way to a new class of devices in that photo-irradiation can be used to alter the electrical properties via strong electron-phonon coupling.

In conclusion, we have investigated the dynamics of time-domain coherent optical phonons in the prototypical transition-metal dichalcogenide, $2H$-MoTe$_{2}$, for different pump fluences.
Both the $E_{1g}$ and $A_{1g}$ modes were observed to possess different decay dynamics.
In particular, the unusually long lifetime coherent $A_{1g}$ phonon mode was observed even in the presence of photo-excited carriers at room temperature and it was found that its decay rate changed significantly with variation in the excitation laser fluence due to weak anharmonic phonon-phonon coupling and fluence-dependent deformation potential electron-phonon interactions.
The present data on ultrafast lattice dynamics can be expected to open up new routes to understand carrier-phonon coupling that govern electronic transport properties in transition-metal dichalcogenides.

This work was supported in part by JSPS KAKENHI Grant Numbers 17H02908 and 25107004. 

\end{document}